\documentclass[twocolumn]{aastex631}

\usepackage{amsfonts}
\usepackage{amssymb}
\usepackage{subfigure}
\usepackage{graphicx}
\usepackage{ulem}
\usepackage{amsmath}

\usepackage{xcolor}

\newcommand{\be}{\begin{equation}}
\newcommand{\ee}{\end{equation}}
\newcommand{\ba}{\begin{eqnarray}}
\newcommand{\ea}{\end{eqnarray}}

\begin{document}

\title{Fuzzy Dark Matter as a Solution to Reconcile the Stellar Mass Density of High-z Massive Galaxies and Reionization History}

\correspondingauthor{Yan Gong}
\email{gongyan@bao.ac.cn}

\author{Yan Gong*}
\affiliation{National Astronomical Observatories, Chinese Academy of Sciences, Beijing 100101, China}

\author{Bin Yue}
\affiliation{National Astronomical Observatories, Chinese Academy of Sciences, Beijing 100101, China}

\author{Ye Cao}
\affiliation{Institute for Frontier in Astronomy and Astrophysics, Beijing Normal University, Beijing, 102206, China}
\affiliation{Department of Astronomy, Beijing Normal University, Beijing 100875, China}

\author{Xuelei Chen}
\affiliation{National Astronomical Observatories, Chinese Academy of Sciences, Beijing 100101, China}
\affiliation{University of Chinese Academy of Sciences, Beijing 100049, China}
\affiliation{Department of Physics, College of Sciences, Northeastern University, Shenyang 110819, China}
\affiliation{Center for High Energy Physics, Peking University, Beijing 100871, China}

\begin{abstract}
The {\it JWST} early release data show unexpected high stellar mass densities of massive galaxies at $7<z<11$, a high star formation efficiency is probably needed to explain this. However, such a high star formation efficiency would greatly increase the number of ionizing photons, which would be in serious conflict with current cosmic microwave background (CMB) and other measurements of cosmic reionization history. To solve this problem, we explore the fuzzy dark matter (FDM), which is composed of ultra-light scalar particles, e.g. ultra-light axions, and calculate its halo mass function and stellar mass density for different axion masses. We find that the FDM model with $m_a\simeq 5\times10^{-23} \rm eV$ and a possible uncertainty range $\sim3\times10^{-23}-10^{-22}\, \rm eV$ can effectively suppress the formation of small halos and galaxies, so that with higher star formation efficiency, both the {\it JWST} data at $z\sim8$ and the reionization history measurements from optical depth of CMB scattering and ionization fraction can be simultaneously matched. We also find that the {\it JWST} data at $z\sim10$ are still too high to fit in this scenario. We note that the estimated mean redshift of the sample may have large uncertainty, that it can be as low as $z\sim9$ depending on adopted spectral energy distribution (SED) templates and photometric-redshift code. Besides, the warm dark matter with $\sim$keV mass can also be an alternative choice, since it should have similar effects on halo formation as the FDM. 
\end{abstract}



\section{Introduction}

Exploration of high-redshift (high-$z$) galaxies can provide valuable information of galaxy formation and history of cosmic reionization. Recently, the $\it James$ $\it Webb$ $\it Space$ $\it Telescope$ ($\it JWST$) has released early observational results of detecting galaxies at $z\gtrsim10$, and some unexpected discoveries have been found that may be in tension with the current galaxy formation theory and the widely accepted $\Lambda$CDM model as well \citep{Boylan-Kolchin22,Lovell22,Mason22,Menci22,Mirocha22,Naidu22a,Naidu22b}. In particular,  \citet{Labbe22} reported a much higher stellar mass density in massive galaxies with $M_*=10^{10}-10^{11}$ $M_{\sun}$ at $7<z<11$, which is more than one magnitude higher at $z\sim 8$,  and three orders of magnitude higher at $z\sim 10$, than the anticipated values from the standard cosmology and star formation scenarios, raising a challenge to the standard cosmological model. 

One way to explain this huge abundance excess is to dramatically enhance the star formation rate (SFR) or UV luminosity function at high-$z$ by invoking a large star formation efficiency $f_*$, which indicates the fraction of baryons that can convert to stars \citep{Boylan-Kolchin22,Lovell22,Mason22,Mirocha22}. Although the typical values of $f_*$ is usually expected to be less than 0.1 based on observations at lower redshifts, it is theoretically possible to have a larger value, even up to $f_* \sim 1$ at the high redshifts, as $f_*$ depends on complicated physics of star formation. Adoption of the higher $f_*$ could partially solve or at least reduce the tension between theories and the $\it JWST$ observation. However, this may raise new tension with the cosmic reionization history, which is closely related to high-$z$ star formation process, and is well constrained by current cosmological observations, such as the cosmic microwave background (CMB) as measured by the $\it WMAP$ and $\it Planck$ satellites \citep[e.g.][]{WMAP9,Planck18}. With a higher star formation efficiency, for a given halo mass $M$, more massive galaxy can be formed,  but then there is an overall increase of ionizing photons produced, and the Universe would be reionized much earlier than as inferred from current observations. 

As only the more massive galaxies is detectable at high redshifts with the current {\it JWST} observations, one possible way to resolve this conflict is to boost the formation of massive galaxies while suppress the formation of small galaxies. One can apply a mass-dependent star formation efficiency $f_*(M)$ to suppress the star formation in small halos, but as we shall discuss later, this may still not be enough to explain the observational results. 

Here we propose and explore the fuzzy dark matter (FDM) as a solution to this problem.
The FDM is a proposed dark matter composed of ultra-light scalar particles \citep{Hu00}, such as ultra-light axions \citep[see e.g.][]{Marsh16a}. Since the FDM particles have extremely low mass with $m_a\sim10^{-22}\ \rm eV$, its de Broglie wavelength can be as large as the size of a dwarf galaxy or even larger. Due to the uncertainty principle in wave mechanics, an effective quantum pressure arises to suppress matter fluctuations below a certain Jeans scale, so small structures cannot form. This will affect the formation and density profile of objects on small scales, and severely suppress the halo mass function at small masses. If the dark matter is composed of FDM, only relatively large halos and massive galaxies are formed, and the rest of dark matter will not cluster, and stay as a smooth component in linear regime. Therefore, it will not significantly affect cosmic reionization history, even when the star formation efficiency is greatly enhanced at high redshifts, since small galaxies do not form in this model.

A number of observations have been performed for probing the FDM, including measurements of density profile and mass function of dwarf galaxies, rotation curves of Milky May, Ly$\alpha$ forest that exploring cosmic structures on small scales, and so on \citep{Schive14,Marsh19,Safarzadeh19,Broadhurst20,Irsic17,Armengaud17,Maleki20}. Possible mass ranges or lower limits of mass of the FDM particles have been derived from these detections. Although some inconsistency may still exist in the current observations, e.g. the observations of Milky Way's dwarf satellites \citep[][]{Safarzadeh19}, an interesting and possible region around $m_a\sim10^{-22}\ \rm eV$ can be located, which is worth for further study.
In the following discussion, we assume a flat $\Lambda$CDM model with $\Omega_b=0.0493$, $\Omega_m=0.3153$, $h=0.6736$, $\sigma_8=0.8111$, $n_s=0.9649$ \citep{Planck18}.

\section{Model}

Since the FDM cannot form small halos below a certain Jeans scale due to quantum pressure, the halo mass function is suppressed at low mass end. The FDM halo mass function has been discussed in previous literatures based on numerical simulations and semi-analytic techniques \citep[see e.g.][]{Schive16,Du18,Schutz20,Niemeyer19}. Here we use a mass-dependent critical overdensity $\delta_{\rm c}(M,z)$ in the analytical mass function to account for this suppression at small mass scale  \citep{Marsh14,Bozek15,Marsh16b,Du17}, which is given by
\be
\delta_{\rm c}(M,z) = \mathcal{G}(M)\, \delta_{\rm crit}(z).
\ee
Here $\delta_{\rm crit}(z)=\delta_{\rm crit}^0/D(z)$, where $\delta^0_{\rm crit}\approx 1.686$ is the critical overdensity for collapse, and $D(z)$ is the linear growth factor normalized at $z=0$. $ \mathcal{G}(M)$ is a factor accounting for mass-dependence, it can be computed with the  {\tt AxionCAMB} code\citep{Hlozek15}, and fitted as functions of halo and axion masses \citep[e.g.][]{Marsh14,Marsh16b,Du17}.

\begin{figure}[t]
\includegraphics[scale = 0.41]{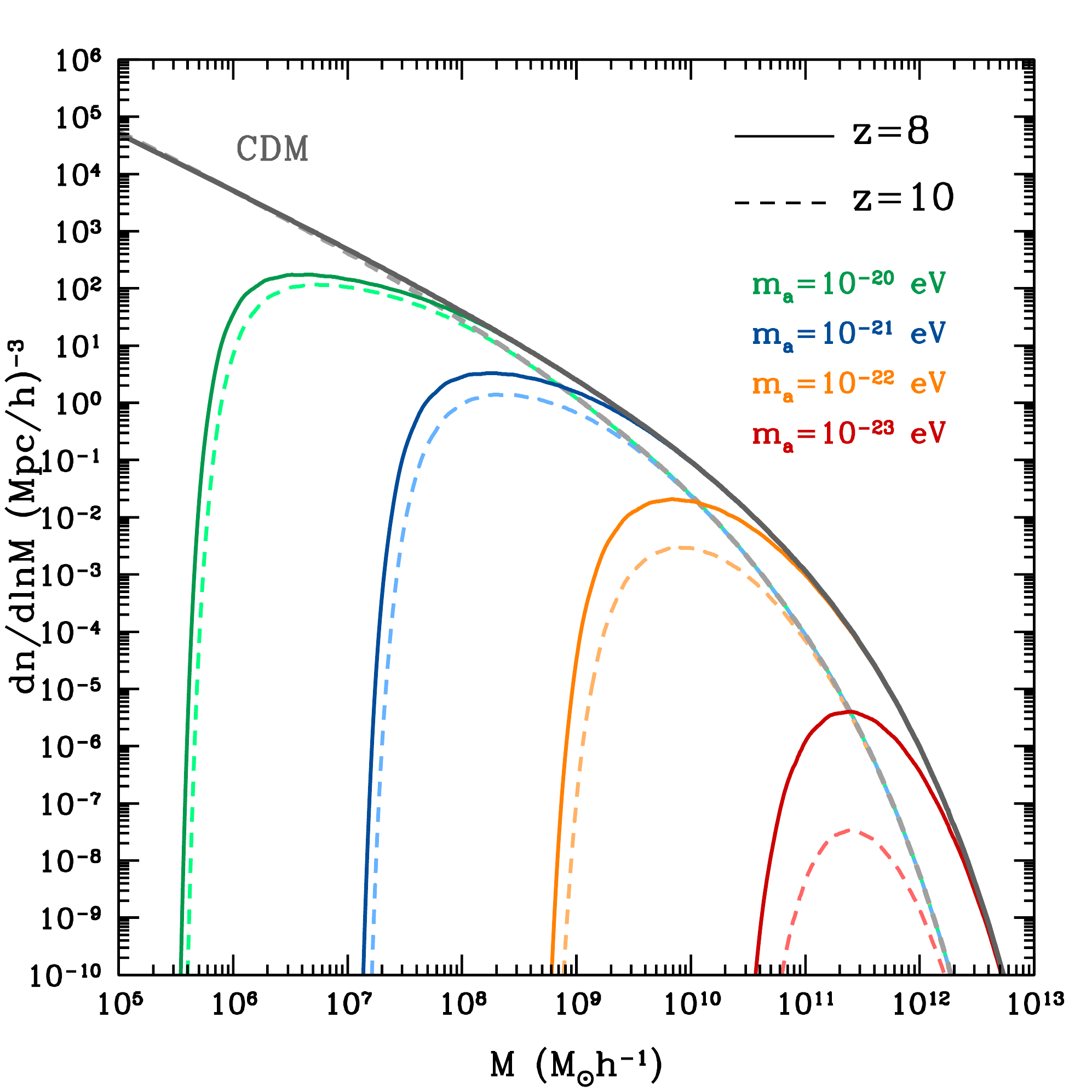}
\caption{The FDM halo mass functions for different axion masses are shown in solid ($z=8$) and dashed ($z=10$) curves. The CDM halo mass functions are also plotted in gray curves for comparison. We can find that small halos with low masses cannot form in FDM.}
\label{fig:dndm}
\end{figure}

The FDM halo mass function can be obtained with this critical density in the Press-Schechter approach \citep{PS74}, 
\be
n(M,z) dM = \frac{\bar{\rho}_m}{M}\nu f(\nu)\frac{d\nu}{\nu},
\ee
where $\bar{\rho}_m=\Omega_m \rho_{\rm crit}$ is the current matter density, $\rho_{\rm crit}$ is the current critical density, and for $\nu f(\nu)$ we take the form as \citep{Sheth99}
\be
\nu f(\nu) = A\sqrt{\frac{\nu'}{2\pi}}\left( 1+\nu'^{-p}\right) e^{-\nu'/2}.
\ee
Here $\nu'=a\nu$, $a=0.707$, $p=0.3$ , $A$ is the normalization factor, $\nu \equiv [\delta_{\rm c}(M,z)/\sigma(M)]^2$, and $\sigma^2(M)$ is the variance of linear power spectrum.
For the FDM linear matter power spectrum, we adopt a numerical result given by \cite{Hu00}. Considering suppression of the FDM power spectrum relative to the CDM case, we have
\be
P_{\rm lin}^{\rm FDM}(k,z) = T^2_{\rm F}(k)\,P_{\rm lin}^{\rm CDM}(k,z).
\ee
Here $P_{\rm lin}^{\rm CDM}$ is the CDM linear power spectrum, which can be estimated analytically \citep{Eisenstein98}. $T_{\rm F}(k)\approx {\rm cos}\,x^3/(1+x^8)$ is the transfer function, and $x=1.61m_{22}^{1/18}k/k_{\rm J}^{\rm eq}$, where $m_{22}=m_a/10^{-22}\ \rm eV$ and $k_{\rm J}^{\rm eq}=9\,m_{22}^{1/2}$ Mpc$^{-1}$ is the comoving Jeans wavenumber scale at epoch of matter-radiation equality. The transfer function has acoustic oscillation features on small scales below $k_{\rm J}^{\rm eq}$, and will approach to unity for large FDM particle mass when $m_a\gg10^{-22}$ eV or $m_{22}\gg1$. 
The FDM halo mass functions at $z=8$ (solid) and $z=10$ (dashed) for different axion masses are shown in Figure~\ref{fig:dndm}. We can see that the abundance of low-mass halos is greatly suppressed in FDM, and only massive halos can form with the same mass function as the CDM case. This allows adopting larger number density for massive galaxies without over-producing ionizing photons.   

\begin{figure*}
\centerline{
\includegraphics[scale=0.3]{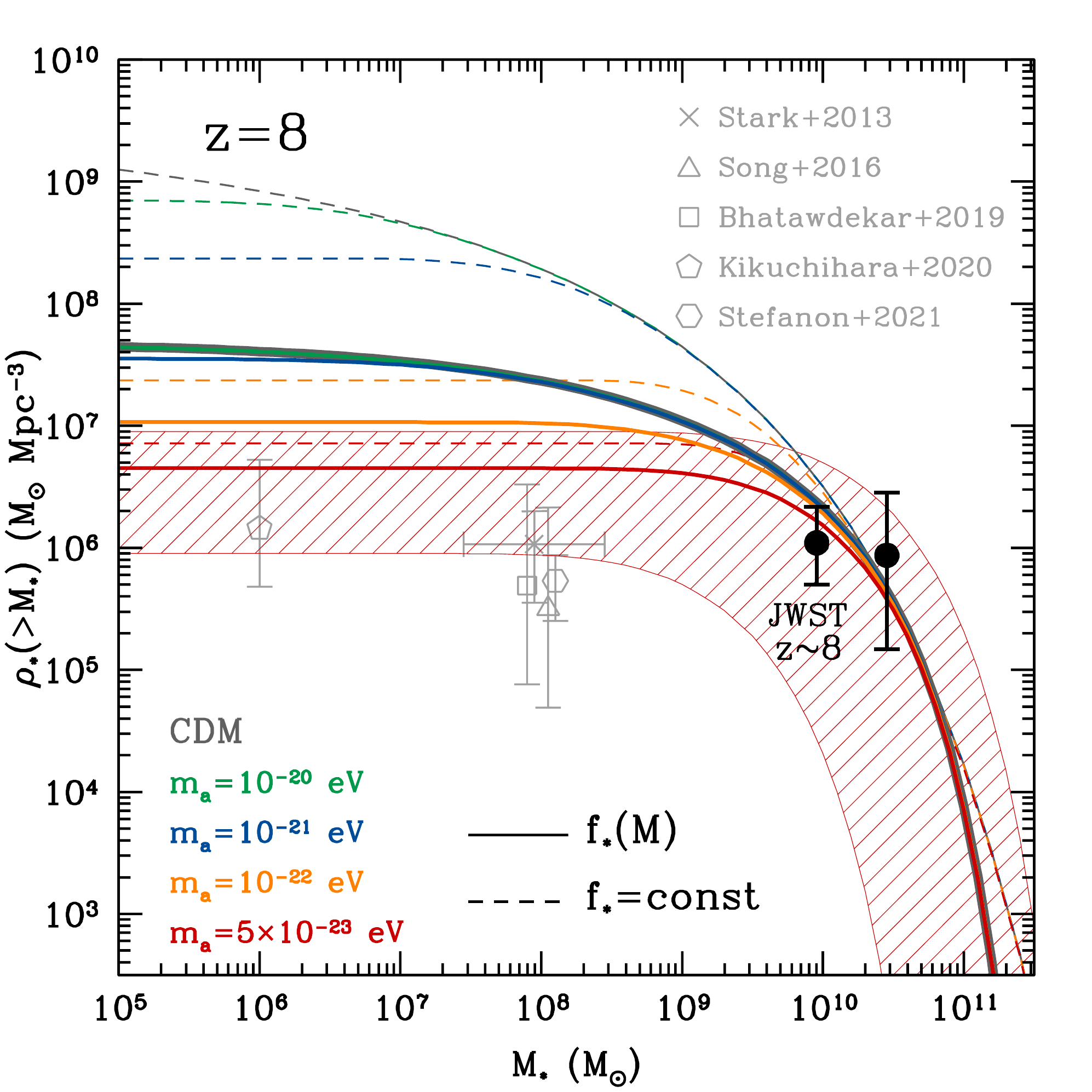}
\includegraphics[scale=0.3]{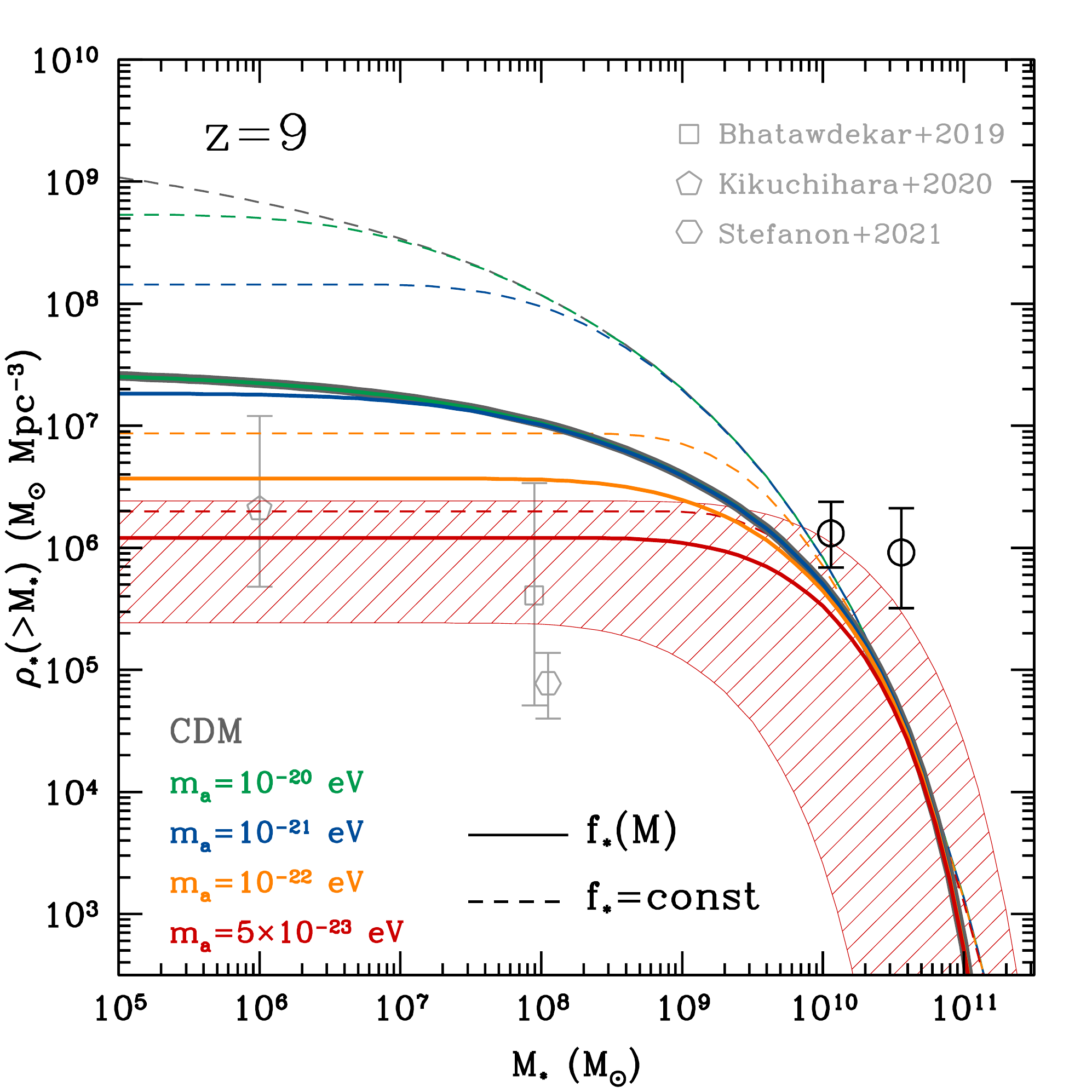}
\includegraphics[scale=0.3]{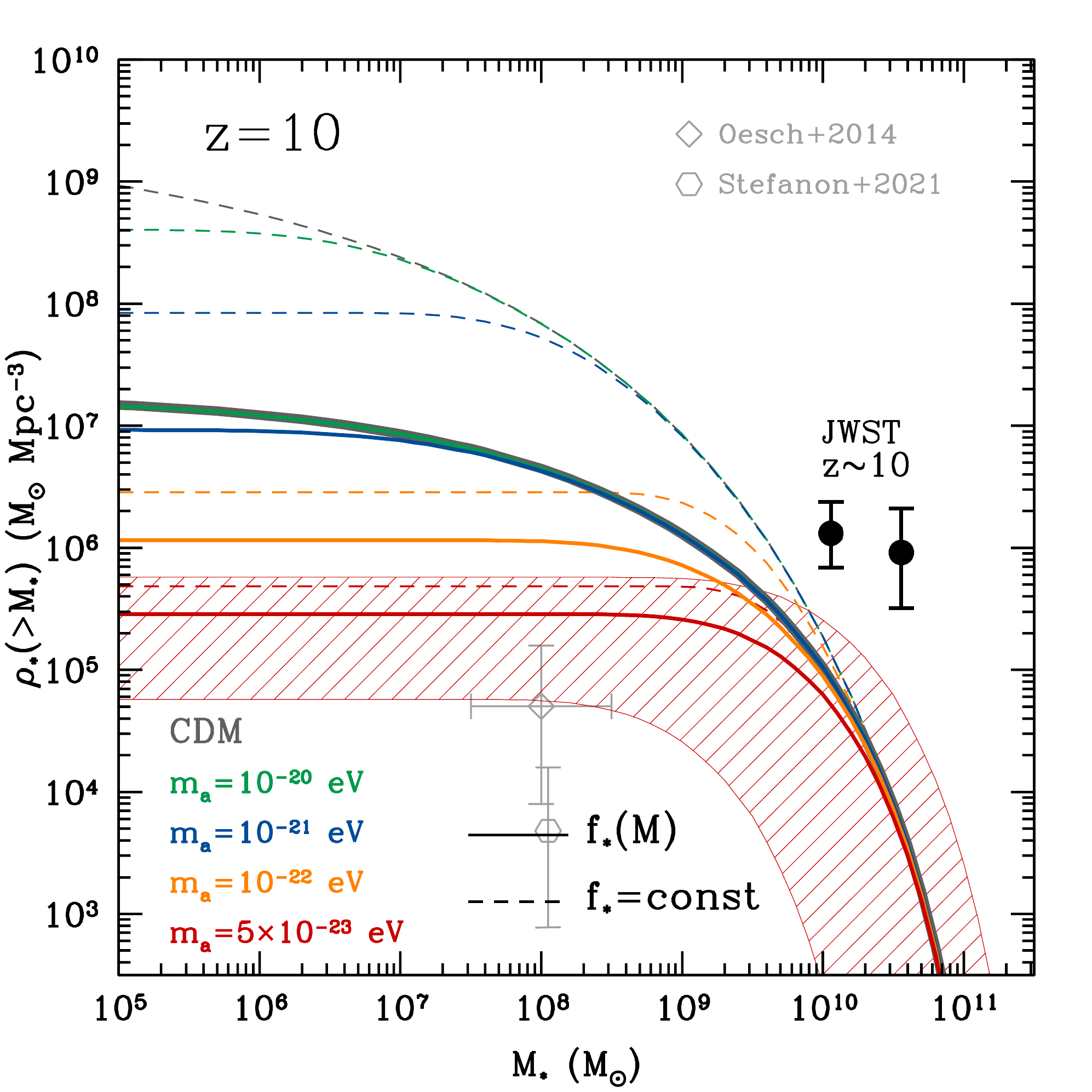}}
\caption{The comoving cumulative stellar mass density as a function of stellar mass for different axion masses at $z=8$ (left panel), $z=9$ (middle panel) and $z=10$ (right panel). The solid curves denote the results derived from the mass-dependent star formation efficiency $f_*(M)$ with $f_*^0=0.5$, and the dashed curves are for constant $f_*=0.5$. The hatched region shows the range of the results of $m_a=5\times10^{-23}\ \rm eV$ with $f_*^0=0.1-1$. For comparison, the results of the CDM case are also shown in gray curves. The {\it JWST} measured values at mean redshift $z\sim8$ and $z\sim10$ as given by  \citet{Labbe22} are over-plotted as black solid dots with error bars in the Left and Right panels, respectively. Our estimate shows that the mean redshift for the second group could be $z\sim9$, so we also plot them as open circles in the Middle panel. The data at low stellar mass are also shown in gray data points for comparison, given by \citet{Stark13}, \citet{Oesch14}, \citet{Song16}, \citet{Bhatawdekar19}, \citet{Kikuchihara20}, and \citet{Stefanon21}. Note that we assume  an intrinsic dispersion of 0.5 dex for the ${\rm log}\,M_{\rm UV}-M_*$ relation when plot the data points given in \citet{Stark13} and \citet{Oesch14}. To show these data clearly, we make little shifts for the data at $M_*=10^8\, M_{\sun}$.}
\label{fig:rho}
\end{figure*}

Hence, the comoving cumulative halo mass density with halo mass greater than $M$ can be calculated by
\be
\rho(>M,z) = \int^\infty_M dM\, M'\, n(M',z).
\ee
Considering the relation between the halo mass and stellar mass, i.e. $M_*=f_*(\Omega_b/\Omega_m)\,M=\epsilon M$, we obtain the cumulative stellar mass density with stellar mass larger than $M_*$,
\be
\rho_*(>M_*,z)=\epsilon\, \rho(>M_*/\epsilon,z).
\ee
We consider two forms of star formation efficiency in this work, i.e. mass-independent $f_*=const$ and mass-dependent $f_*(M)$, for the latter we assume a double power-law (DPL) form as given in \citet{Mirocha2017}
\be
f_*(M) = \frac{2f_{*}^0}{\left(\frac{M}{M_{\rm p}}\right)^{\alpha_{\rm lo}}+\left(\frac{M}{M_{\rm p}}\right)^{\alpha_{\rm hi}}},
\ee
where we adopt $M_{\rm p}=2.8\times10^{11}~M_\odot$,  $\alpha_{\rm lo}=0.49$, and $\alpha_{\rm hi}=-0.61$.  These parameters are calibrated to match the observed high-$z$ LFs \citep{Bouwens2015a}. $f_*^0$ is the star formation efficiency at halo mass $M_{\rm p}$, and they find $f_*^0=0.025$ \footnote{Note that there is an additional factor 2 in the numerator compared to that given in \citet{Mirocha2017}.}. We will adjust $f_*^0$ and $f_*$ to match the $\it JWST$ data in this work.

As we mentioned, a large star formation efficiency can significantly affect the cosmic reionization history, and may violate the current measurements of epoch of reionization. To evaluate this effect, we can first calculate the hydrogen volume filling factor $Q_{\rm HII}(z)$ as a function of redshift.

The evolution of $Q_{\rm HII}$ follows \citep{Wyithe2003,Madau99}
\begin{equation}
\frac{dQ_{\rm HII}}{dt}=f_{\rm esc} \frac{\dot{n}_{\rm ion}}{\bar{n}_{\rm H}}-C_{\rm HII}(z)\alpha_B(T_{\rm HII}) \bar{n}_{\rm H}(1+z)^3x_e,
\end{equation}
where $f_{\rm esc}$ is the escape fraction set to be 0.1 \citep{Sun2021},  $\bar{n}_{\rm H}$ is the mean number density of hydrogen (both neutral and ionized) atoms at present Universe, $C_{\rm HII}=3.0$ is the clumping factor of the ionized gas \citep{Kaurov2015}, $\alpha_B$ is the Case B recombination coefficient, and $T_{\rm HII}$ is the kinetic temperature. We always take $T_{\rm HII}=2\times 10^4$ K \citep{Robertson2015}, so that $\alpha_B=2.5\times10^{-13}$ cm$^3$s$^{-1}$. Here for simplicity we assume the helium has the same first stage ionization (i.e. He II) fraction as hydrogen (the full ionization to He III would be much later), so the total ionization fraction can be written as
\begin{equation}
x_e=Q_{\rm HII}(1+\frac{Y_{\rm He}}{4}),
\end{equation}
where $Y_{\rm He}=0.25$ is the Helium element abundance. For the emission rate of ionizing photons per unit comoving volume $\dot{n}_{\rm  ion}$, we take
\begin{equation}
\dot{n}_{\rm ion}=N_{\rm ion} \frac{\Omega_b}{\Omega_m} \frac{1}{t_{\rm  SF}(z)} \int_{M_{\rm min}}^\infty dM\, n(M,z)\,f_*(M)M ,
\end{equation}
where $N_{\rm ion}\approx 4000$ is the total ionizing photons produced per stellar baryon throughout its lifetime for typical Pop II galaxies (e.g. see {\tt Starburst99}\footnote{\url{https://www.stsci.edu/science/starburst99/docs/default.htm}},\citealt{Leitherer1999,Vazquez2005,Leitherer2010,Leitherer2014}), $t_{\rm SF}$ is the star-forming timescale, and we assume that it equals to 10\% of the Hubble time at redshift $z$ \citep{Wyithe06,Lidz11}. $M_{\rm min}$ is the minimum halo mass corresponding to a virial temperature of $10^4$ K, halos above this mass can sustain effective cooling via the Ly$\alpha$ transition \citep{Barkana2001}. We find that, for example, $M_{\rm min}=4.6\times10^7$, $8.0\times10^7$, and $2.0\times10^8\, M_{\sun}$ at $z=15$, 10, and 5, respectively. The major contribution of ionizing photons depends on the shape of $f_*(M)n(M,z)$, and basically it is dominated by low-mass galaxies in the CDM model. In the FDM model, it can be dominated by massive galaxies, since small galaxies can barely form when $m_a$ is small.

The optical depth of the CMB scattering is adopted as a quantity to characterize the cosmic reionization history, which can be estimated by
\begin{equation}
\tau=\int_0^{\infty} \sigma_{\rm T} \bar{n}_{\rm H}(1+z)^3x_e\frac{cdz}{(1+z)H(z)},
\end{equation}
where $\sigma_{\rm T}=6.65\times10^{-25}$ cm$^{-2}$ is the Thompson scattering cross-section.
We integrate up to $z_{\rm max}=30$ since the reionization is negligible before this redshift in our model.

\section{Result and Discussion}

\begin{figure*}
\centerline{
\includegraphics[scale = 0.4]{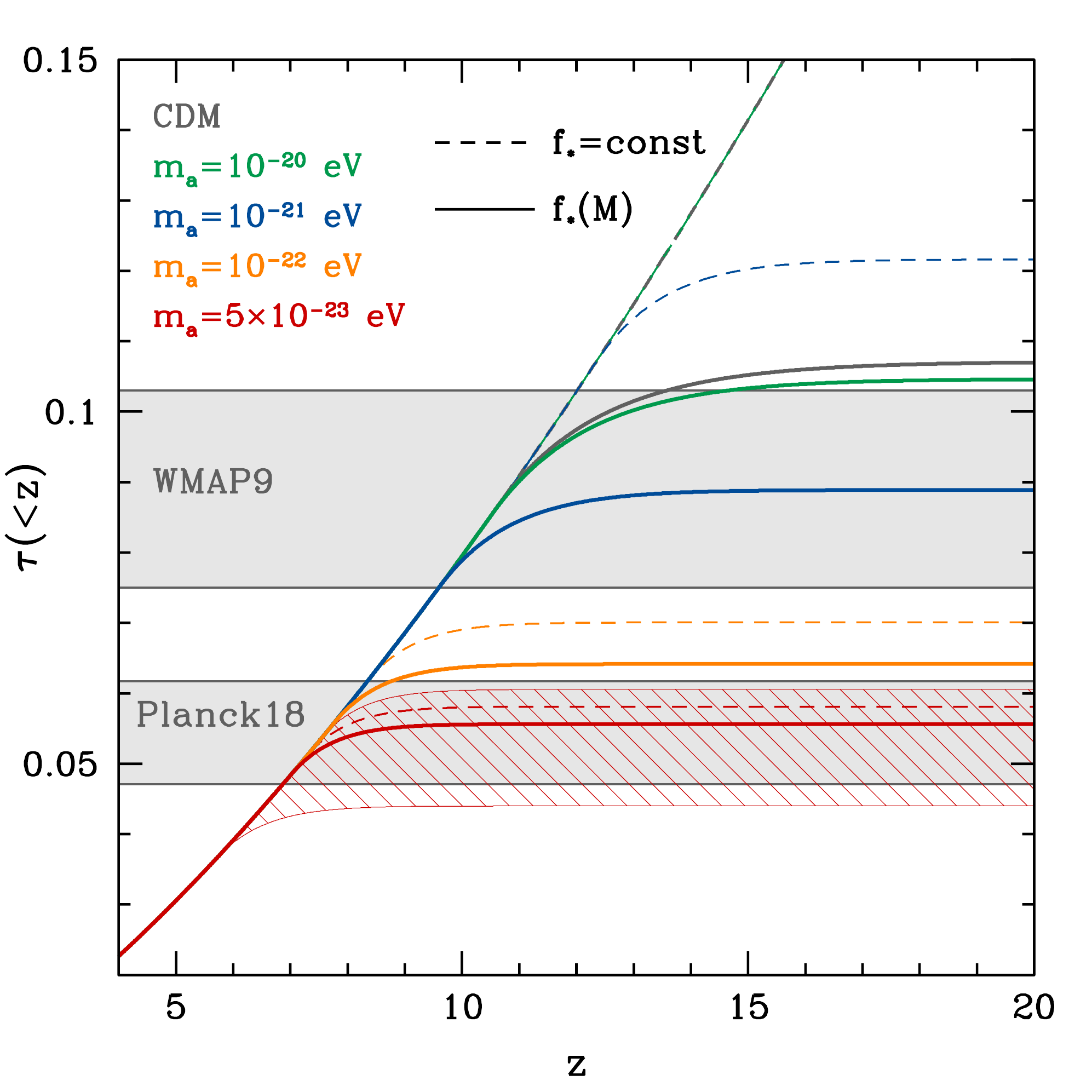}
\includegraphics[scale=0.4]{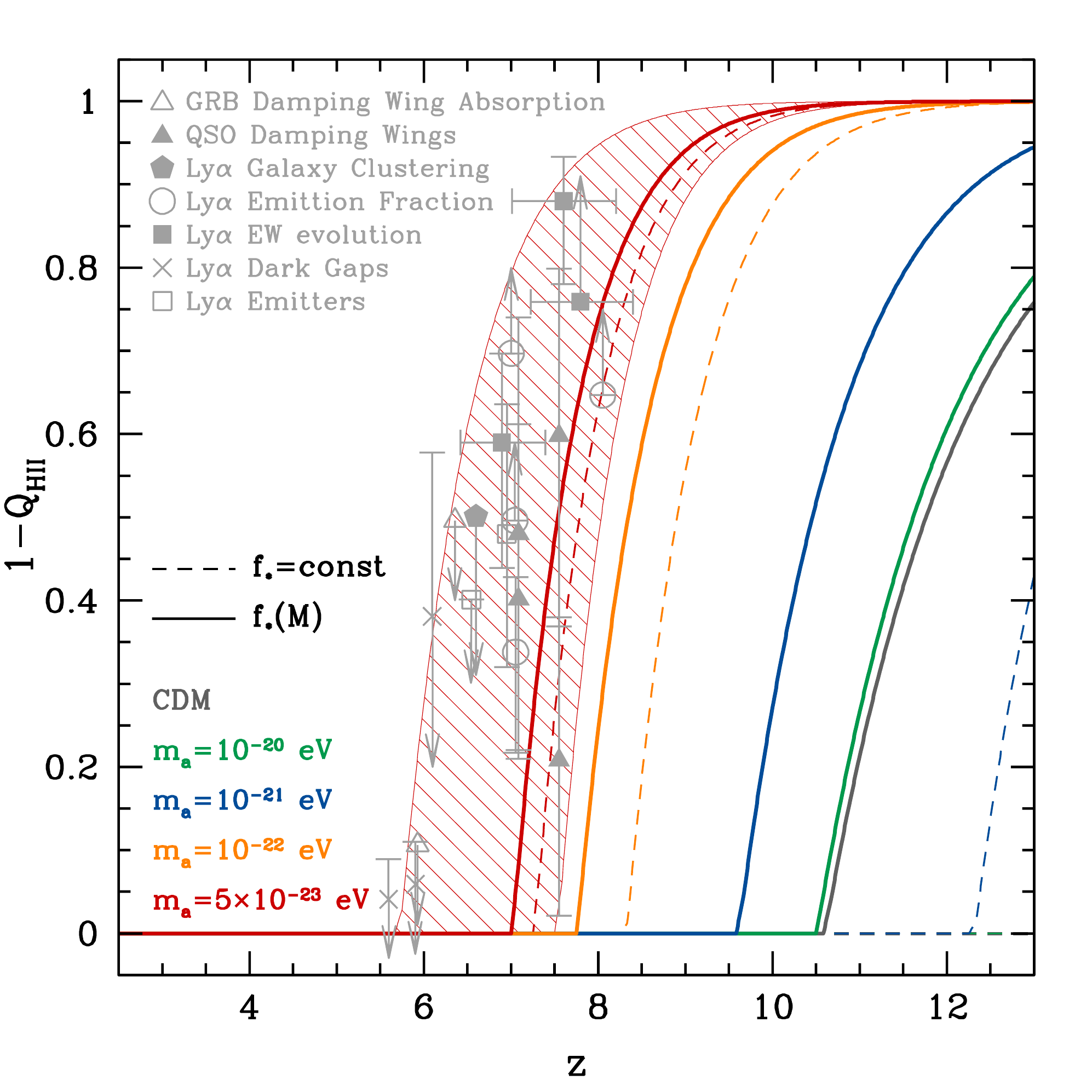}}
\caption{{\it Left panel:} The optical depth $\tau$ as a function of redshift for the FDM and CDM cases. The solid and dashed curves denote the results from mass-dependent $f_*(M)$ with $f_*^0=0.5$ and constant $f_*=0.5$, respectively. The hatched region is for the results derived from $f_*^0=0.1-1$.The gray shaded regions are the $1\sigma$ (68.3\% C.L.) results given by WMAP9 \citep{WMAP9} and Planck18 \citep{Planck18}. {\it Right panel:} The average IGM neutral hydrogen fraction characterized by $1-Q_{\rm HII}$ as a function of redshift. The gray data points show the results of different measurements in previous studies \citep[see e.g.][and the references therein]{Robertson2015,Bouwens2015b,Mason19}.}
\label{fig:tau}
\end{figure*}

In Figure~\ref{fig:rho}, we show the comoving cumulative stellar mass density $\rho_*(>M_*)$ as a function of $M_*$ at $z=8$ (Left panel), 9 (Middle panel) and 10 (Right panel) for the FDM and CDM cases. As can be seen, $\rho_*(>M_*)$ is effectively suppressed at small stellar masses compared to the CDM case, while at the higher mass end the FDM and CDM curves are identical. For example, at $z=8$, assuming the mass-dependent star formation efficiency with $f_*^0=0.5$, we find that $\rho_*(>M_*)$ almost becomes flat when $M_*\lesssim10^{6}$, $10^7$, $10^8$, and $10^9\ M_{\sun}$,  when $m_a=10^{-20}$, $10^{-21}$, $10^{-22}$, and $5\times10^{-23}\ \rm eV$, respectively. As expected, this effect is more significant in the mass-dependent $f_*(M)$ case than the constant $f_*$ case, since $f_*(M)$ would become smaller at small $M_*$. \citet{Labbe22} has given the estimates in two redshift groups ($z\sim8$ and $z\sim10$), and at each redshift for $M_*>10^{10} M_\odot$ and  $M_*>10^{10.5} M_\odot$. We also over-plotted these estimates in black solid dots in Figure~\ref{fig:rho}. 

If we assume the maximum $f_*$ or $f_*^0$ can reach to 1, i.e. all baryons will convert to stars in the constant $f_*$ case or at $M_{\rm p}$ in the $f_*(M)$ case, it seems that both CDM and FDM models considered here can match the data measured by $\it JWST$ at $z\sim8$ (refer to the hatched region in the left panel of Figure~\ref{fig:rho}). However, as we will discuss below, most of them will not be consistent with the measurements of cosmic reionization history. 

On the other hand, we can see that at $z\sim10$ ( the Right panel of Figure~\ref{fig:rho}), none of the CDM or FDM model we considered can fit the {\it JWST} data even assuming $f_*$ or $f_*^0=1$, as have been found in studies assuming the $\Lambda$CDM model \citep[e.g.][]{Boylan-Kolchin22,Menci22}. This may be due either to strong deviation of cosmological evolution from the $\Lambda$CDM model,  or due to issues of galaxy selection, measurements of galaxy stellar mass and redshift, dust extinction, and sample variance \citep{Endsley22,Ferrara22,Ziparo22,Adams23}.  

In particular, the stellar mass and redshift of these galaxies are estimated photometrically in \cite{Labbe22} using the  {\tt EAZY} code \citep{Brammer08}. Based on other studies, the stellar mass can be even one order of magnitude lower with different assumptions of SED template fitting \citep[e.g.][]{Endsley22}, and then $\rho_*(>M_*)$ also will become lower accordingly. If this is true, as shown in Figure~\ref{fig:rho}, the tension between the JWST data and the CDM model with small $f_*^0<0.1$ can be relaxed at $z\sim8$. However, even so the JWST data at $z\sim10$ seem still too high for the CDM model, although they can match our FDM model with large $f_*^0>0.1$ (red hatched region). In addition, the photometric redshift estimation may also have larger errors, especially at these very high redshifts \citep{Adams23}. To make a simple and practical assessment of this uncertainty, we use another widely used photo-$z$ code, i.e. $\tt LePhare$ \citep{Arnouts99,Ilbert06}, to derive the photometric redshifts of the seven galaxies given in \cite{Labbe22}. A quick check shows that the mean photo-$z$ of the three galaxies in $7<z<9$ and the four galaxies in $9<z<11$ are $\bar{z}_p=8.6\pm0.3$ and $9.4^{+0.6}_{-0.4}$, respectively, compared to $\bar{z}_p=8.3$ and $10.0$ given by \cite{Labbe22}. Our result is also consistent with other works, e.g. \citet{Bouwens22} gives $\bar{z}_p=9.0^{+0.7}_{-0.6}$ for $\bar{z}_p=10$ in \cite{Labbe22}. This indicates that there could be relatively large uncertainty in the current galaxy redshift estimation, which depends on the selected photo-$z$ analysis code and spectral energy distribution (SED) templates. If we shift the density mean redshift from $z\sim10$  to $z\sim9$, as shown in the middle panel of Figure~\ref{fig:rho},  the data could be explained by the current models within 1$\sigma$ error.

\begin{figure*}
\centerline{
\includegraphics[scale=0.3]{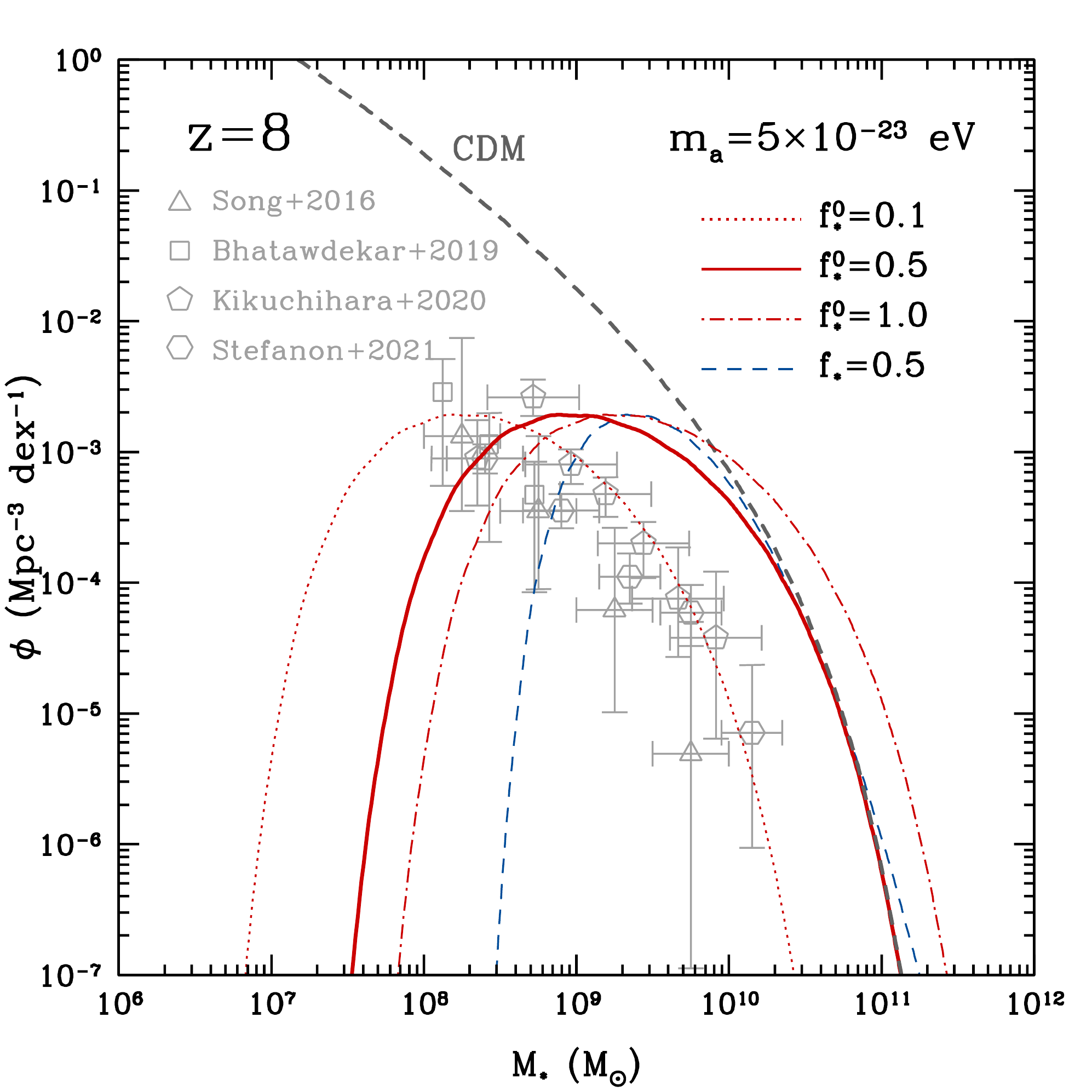}
\includegraphics[scale=0.3]{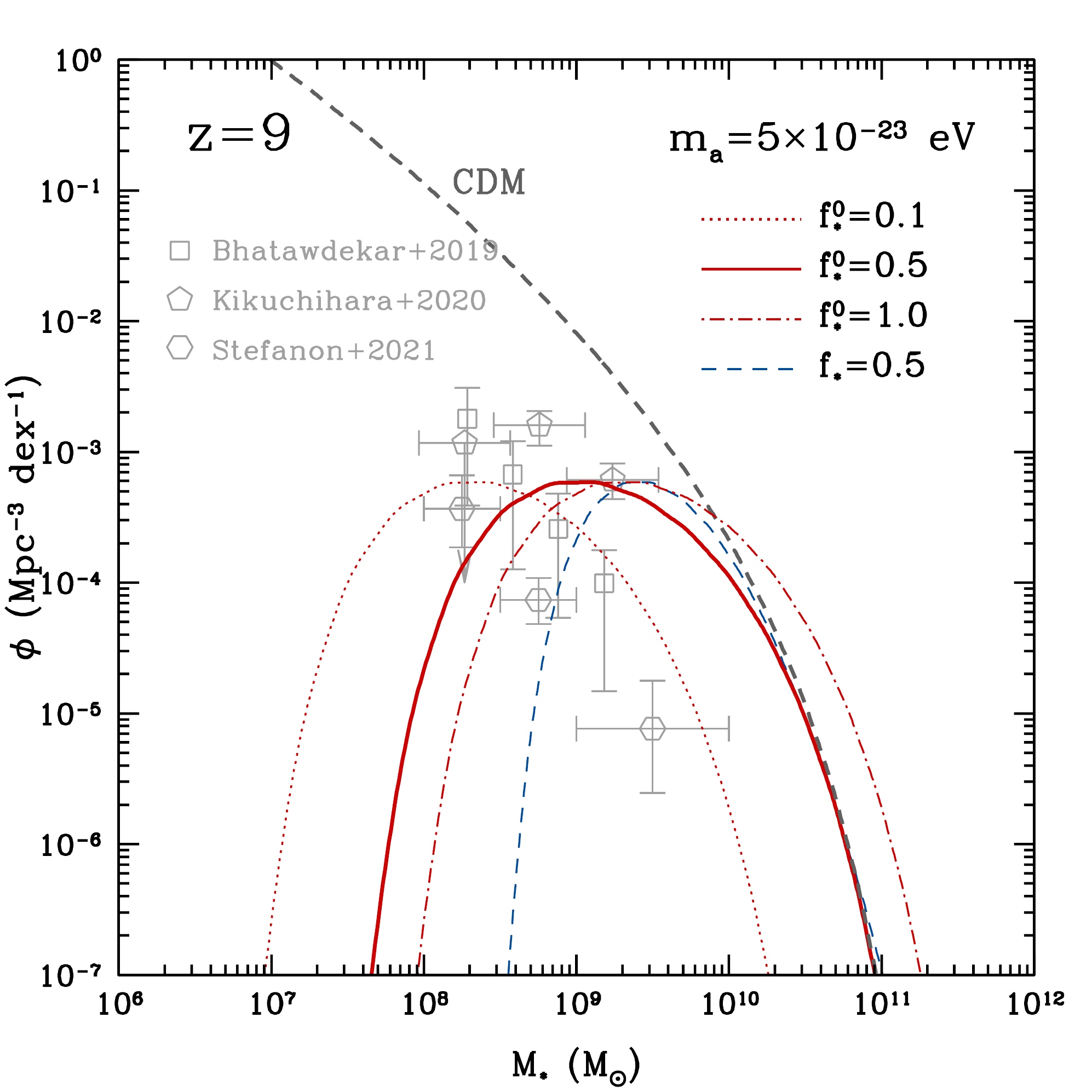}
\includegraphics[scale=0.3]{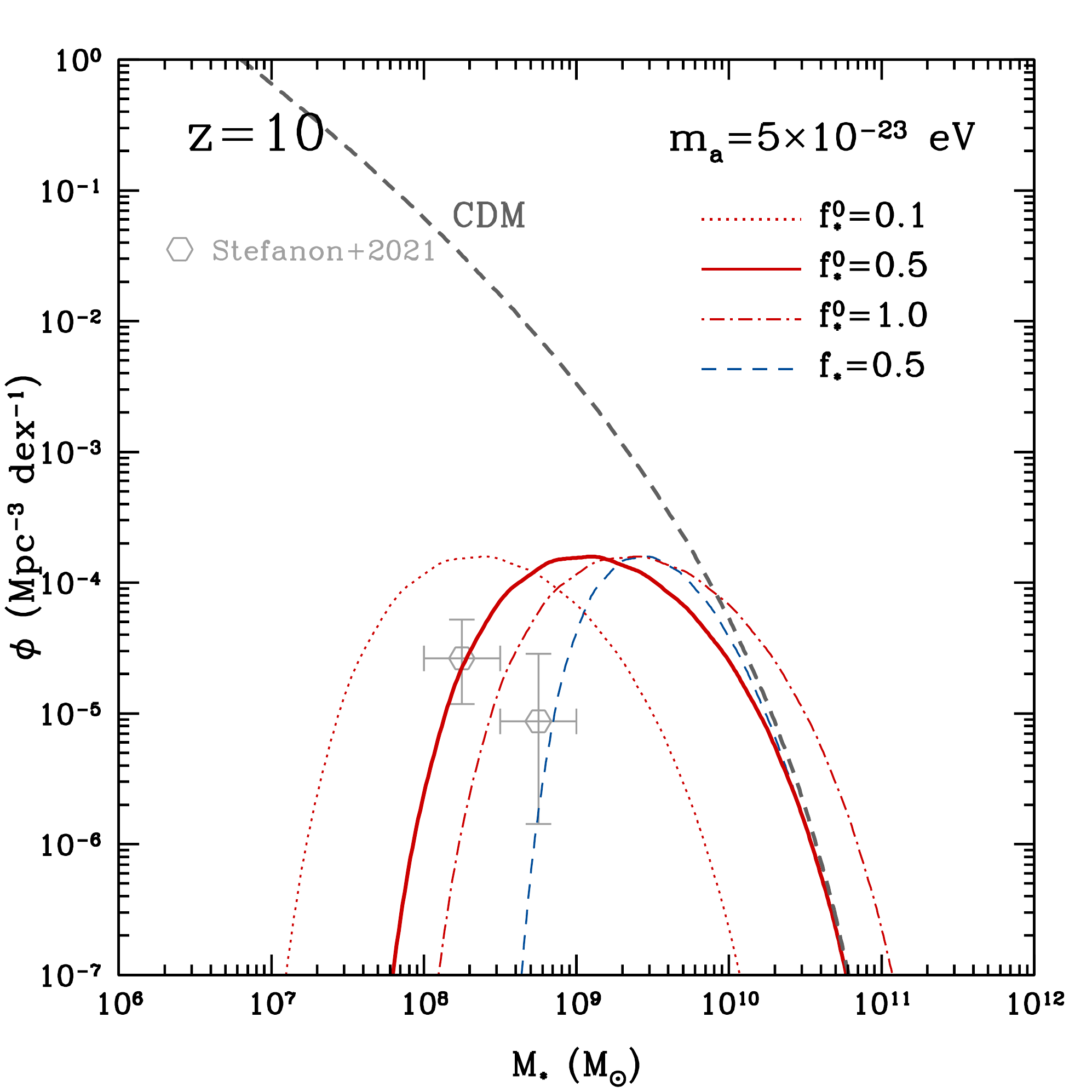}}
\caption{The GSMFs derived from the FDM halo mass functions with $m_a=5\times10^{-23}$ eV and different star formation efficiencies at $z=8$, 9, and 10. The red dotted, solid, and dash-dotted curves denote the results with $f_*^0=0.1$, 0.5 and 1, respectively. The result from the CDM halo mass function is also shown in gray dashed curve for comparison. The blue dashed curve is the GSMF with a constant star formation efficiency $f_*=0.5$. The observational data at $M_*\simeq10^8-10^{10}$ $M_{\sun}$ from HST and other previous measurements are shown as gray data points \citep{Song16,Bhatawdekar19,Kikuchihara20,Stefanon21}.}
\label{fig:GSMF}
\end{figure*}

Besides, for comparison, we also show the data of stellar mass density at low stellar mass end down to $M_*\sim 10^6$ and $10^8\, M_{\sun}$, which are given by previous studies \citep{Stark13, Oesch14, Song16, Bhatawdekar19, Kikuchihara20, Stefanon21}.  We can see that, although some data points are lower and show discrepancy compared to the JWST data at high stellar mass, our model with $m_a=5\times10^{-23}\, \rm eV$ and $f_*^0=0.1-1$ (red hatched region) is consistent with most of these data within 1$\sigma$ confidence level. This means that our FDM model has the potential to explain both the measurement of cumulative stellar mass density from {\it JWST} at high stellar mass and the data obtained at low stellar mass. One the other hand, it seems that the CDM model can hardly fit all these data in the mean time, no matter assuming large or small star formation efficiency.

In the left panel of Figure~\ref{fig:tau}, we plot the optical depth $\tau$ as a function of $z$ for the CDM and FDM models. The $68.3\%$ confidence level (C.L.) results of the 9-year $\it WMAP$ (WMAP9) and $\it Planck$ 2018 (Planck18) \citep{WMAP9,Planck18} are also shown in gray parallel bands, which give $\tau=0.089\pm0.014$ and $0.0540\pm0.0074$, respectively.  Note that the WMAP9 result is actually consistent with $\it Planck$ 2015 result with $\tau=0.079\pm0.017$ \citep{Planck15} in $1\sigma$ error. We can find that the case of $m_a=5\times10^{-23}\ \rm eV$ can fit the Planck18 result very well for both mass-dependent and constant star formation efficiencies, and the $m_a=10^{-21}\ \rm eV$ case is consistent with the WMAP9 results for mass-dependent $f_*(M)$. Both the CDM cases of assuming mass-dependent $f_*(M)$ with $f_*^0=0.5$ and constant $f_*=0.5$ (gray curves) predict much higher $\tau$, that cannot fit the measurements. Hence, although most of these FDM and CDM models can fit the $\it JWST$ data at $z\sim8$ (see the left panel of Figure~\ref{fig:rho}), only $m_a=5\times10^{-23}$ and $10^{-21}\ \rm eV$ can give good match to  the cosmic reionization history measured by $\it Planck$ and $\it WMAP$ satellites, respectively.

We also show the neutral hydrogen fraction of the intergalactic medium (IGM) characterized by $1-Q_{\rm HII}$ as a function of redshift in the right panel of Figure~\ref{fig:tau}. The results from Ly$\alpha$ galaxies, gamma-ray burst (GRB), and quasi-stellar object (QSO) measurements are shown in gray data points \citep[see e.g.][]{Robertson2015,Bouwens2015b,Mason19}. We can see that the FDM model with $m_a=5\times10^{-23}\, \rm eV$ and $f_*^0=0.1-1$ (red hatched region) are in good agreement with these data, and the models with other axion masses cannot fit the data very well. Hence, it indicates that the FDM model with $m_a=5\times10^{-23}\, \rm eV$ can match all the data of high-z galaxy stellar mass density and reionization history measured by optical depth and IGM neutral hydrogen fraction.

We should note that there could be large uncertainties in the current result.  On one hand, as mentioned, the current $\it JWST$ stellar mass density data may still have large errors that needs to be further studied. This can directly affect the analysis and the result of FDM mass determination. For instance, if the stellar mass is overestimated in \cite{Labbe22}, the FDM particle mass will be larger and the star formation efficiency can be smaller. On the other hand, large uncertainties of the model and parameters could exist in our analytical estimation,  e.g. star formation efficiency, escape fraction, clumping factor, etc. \citep{Finkelstein19,Yung20}. So the FDM particle mass we derive should have an uncertainty range. Considering the uncertainties in the measurements of stellar mass density, reionization history, and the model and parameters, we can find a possible $m_a$ range of $\sim 3\times10^{-23}-10^{-22}\, \rm eV$.

In addition, as we mentioned above, it shows some discrepancy between the $\it JWST$ cumulative stellar mass density data at high stellar mass and the previous data obtained by Hubble space telescope (HST) and other telescopes at low stellar mass in Figure~\ref{fig:rho}, especially at high redshift. To further check this problem and compare with our FDM model, we calculate the galaxy stellar mass functions (GSMFs) from the FDM halo mass functions with $m_a=5\times10^{-23}\, \rm eV$ and different star formation efficiencies at $z=8$, 9 and 10, respectively, and we show the results in Figure~\ref{fig:GSMF}. For comparison, we also show the observational data at low stellar masses given in previous measurements by HST and Spitzer \citep{Song16,Bhatawdekar19,Kikuchihara20,Stefanon21}. We only plot the data at $M_*\gtrsim10^8\, M_{\sun}$ here, since the data at  $M_*< 10^8\, M_{\sun}$ usually have large uncertainties on both stellar mass (that can be higher than $10^8\, M_{\sun}$) and galaxy volume density at $z\gtrsim8$ \citep[see e.g.][]{Kikuchihara20,Stefanon21}. 

We can find that the FDM model with $m_a=5\times10^{-23}\, \rm eV$ and $f_*^0\sim0.1$ can fit most of the GSMF data at $z=8$ and 9,  and it seems that the data prefer a even lower star formation efficiency with $f_*^0<0.1$ at $z=10$. This is obviously not in agreement with the result derived from the $\it JWST$ cumulative stellar mass density data at high stellar mass, which prefer a large star formation efficiency with $f_*^0\gtrsim0.5$ as shown in Figure~\ref{fig:rho}. This implies that there is disagreement between $\it JWST$ data given in \citet{Labbe22} and the previous measurements at low stellar mass. So further studies are needed to confirm the stellar mass and redshift in the current $\it JWST$ data, and spectroscopic observations should be especially helpful by providing the SEDs of high-$z$ galaxies.


\section{Conclusion}

We explore the FDM as a solution to reconcile the unexpected high stellar mass density of massive galaxies at $7<z<11$ obtained in the $\it JWST$ early release measurements and the reionization history. To explain this high density, a large star formation efficiency is probably needed, which may greatly boost the number of ionizing photons and violate the cosmic reionization history measured by current CMB and other observations. The FDM that is composed of ultra-light scalar particles, e.g. ultra-light axions, can effectively suppress the formation of small halos and galaxies due to the galaxy-size de Broglie wavelength. This provides a possible way to solve this problem.

By exploring the FDM with different axion masses, we find that the FDM model can simultaneously fit the cumulative stellar mass density data from the $\it JWST$ at $z\sim8$ and the optical depth of the CMB scattering $\tau$, when the axion mass $m_a\simeq5\times10^{-23}$ and $10^{-21}\ \rm eV$ for the Planck18 and WMAP9 results, respectively. After considering the reionization history measurements by the IGM ionization fraction $Q_{\rm HII}$, only $m_a\simeq5\times10^{-23}\, \rm eV$ with $f_*^0=0.1-1$ is preferred. Although the $\it JWST$ stellar mass density data at $z\sim10$ are still too high to explain, we find that the current photo-$z$ estimation may have large uncertainties, and the mean redshift of the sample can be as low as $z\sim9$ if using different SED templates and photo-$z$ codes. Besides, other terms, such as galaxy selection, uncertainty of stellar mass, dust extinction and sample variance, can also affect the results. By comparing with the GSMF data given by previous measurements, we find large disagreement with the current $\it JWST$ data, which indicates that further studies are needed, especially the measurements by spectroscopic observations. With the current uncertainties from both the observational data and model considered, we can estimate a possible $m_a$ range from $\sim3\times10^{-23}$ to $\sim10^{-22}\, \rm eV$. We also notice that, in addition to the FDM, the warm dark matter with $\sim$keV mass could have similar effect on halo formation as the FDM, and should be worth to investigate in the future work.

\begin{acknowledgments}
We acknowledge the support of National Key R\&D Program of China No.2022YFF0503404, MOST-2018YFE0120800, 2020SKA0110402, NSFC-11822305, NSFC-11773031, and NSFC-11633004, the Chinese Academy of Science grants QYZDJ-SSW-SLH017, XDB 23040100, XDA15020200. This work is also supported by the science research grants from the China Manned Space Project with NO.CMS-CSST-2021-B01 and CMS- CSST-2021-A01.
\end{acknowledgments}


\end{document}